# Solution Processed Infrared- and Thermo- Photovoltaics based on 0.7 eV Bandgap PbS Colloidal Quantum Dots

Yu Bi,[a] Arnau Bertran,[a] Shuchi Gupta,[a] Iñigo Ramiro,[a] Santanu Pradhan,[a] Sotirios Christodoulou,[a] Shanmukh-Naidu Majji,[a] Mehmet Zafer Akgul [a] and Gerasimos Konstantatos [a, b] *

Harnessing low energy photons is of paramount importance for multi-junction high efficiency solar cells as well as for thermo-photovoltaic applications. However, semiconductor absorbers with bandgap lower than 0.8 eV have been limited to III-V (InGaAs) or IV (Ge) semiconductors that are characterized by high manufacturing costs and complicated lattice matching requirements in their growth and integration with the higher bandgap cells. Here, we have developed solution processed low bandgap photovoltaic devices based on PbS colloidal quantum dots (CQDs) with a bandgap of 0.7 eV suited for both thermo-photovoltaic as well as low energy solar photon harvesting. By matching the spectral response of those cells to that of the infrared solar spectrum, we report a record high short circuit current ($J_{SC}$) of 37 mA/cm² under full solar spectrum and 5.5 mA/cm2 when placed at the back of a silicon wafer resulting in power conversion efficiencies (PCE) of 6.4 % and 0.7 % respectively. Moreover, the device reached an above bandgap PCE of ~6 % as a thermo-photovoltaic cell recorded under a 1000 °C blackbody radiator.

## Introduction

The development of infrared photovoltaics is of paramount importance for reaching high efficiency solar cells. Multi-junction solar cells comprising a stack of two or more solar absorbers with varied bandgaps on top of each other have been extensively explored as an effective way to surpass the single junction cell Shockley-Queisser efficiency limit.[1, 2] Theoretical calculations have shown that absorbers with low bandgap around 0.7 eV for the rear subcells are required, in combination with front-cell absorbers of 1.16 eV and 1.86 eV bandgap in a triple junction cell, to reach efficiency above 49 % under one sun conditions.[1] In addition to that and beyond solar harnessing, low bandgap semiconductors (0.5 eV-0.74 eV) can be used in thermo-photovoltaic (TPV) applications to harness thermal energy or facilitate waste heat recovery.[3-6] However, the availability of low bandgap semiconductors is limited especially taking into account cost considerations. Potential candidates, considered hitherto, as low bandgap absorber materials are GaSb (0.73 eV),[7-9] Ge (0.67 eV),[10, 11] InGaAs (0.36-0.75 eV),[10, 12, 13], InGaAsSb (0.5-0.6 eV).[3, 14] Yet the growth of these materials often requires costly high ultra-high vacuum facilities and lattice matched substrates. **Figure 1a** (left axis) plots the solar power spectrum. It is evident that most of the standard technologies at hand (such as perovskites, polymer and Si solar cells) are capable of harnessing only a portion of it typically up to 1100 nm (marked in blue line). However, as shown by the integrated $J_{SC}$ curve of **Figure 1a** (right axis) overlaid with the most commonly used technologies currently in place, a significant number of photons are available for harnessing beyond their absorption edge. Colloidal quantum dots (CQD) offer the advantages of low-cost, solution processability, wide range bandgap tunability, and therefore present an excellent candidate for low band gap absorbers.[15-17] **Figure 1b** illustrates the expected additional current that could be gained when using a PbS CQD solar cell in a 4-terminal tandem configuration with some representative available technologies of different bandgaps, as function of the PbS CQD bandgap (1.28, 0.97 and 0.76 eV). The expected additional current was calculated by using optical modeling based on transfer matrix method (TMM).[18-20] The device structure used in the simulation is as follows: ITO (80 nm)/ZnO (40 nm)/PbS (100-600 nm)/Au (80 nm). Refractive index n and extinction coefficient k of different layers used in the simulation were measured by a spectroscopic ellipsometer. The optimized PbS active layer thickness was determined by the maximum $J_{SC}$ obtained from the TMM simulation, leading to the corresponding optimized EQE curve. The corresponding EQE curves of the 1.28, 0.97 and 0.76 eV PbS CQDs are plotted in **Figure S1**. **Figure 1c** further plots the additional power points that could then be reached considering $V_{OC}$ and FF values typically reported for PbS CQD solar cells.[21, 22] ($V_{OC}$ values of 0.7 V, 0.41 V and 0.31 V were assumed for the 1.28 eV, 0.97 eV and 0.76 eV bandgap PbS

[a.] ICFO—Institut de Ciencies Fotoniques, The Barcelona Institute of Science and Technology, Av. Carl Friedrich Gauss, 3, 08860 Castelldefels (Barcelona), Spain
[b.] ICREA—Institució Catalana de Recerca i Estudis Avançats, Passeig Lluís Companys 23, 08010 Barcelona, Spain
* Corresponding Author: Gerasimos.Konstantatos@icfo.e

†Electronic Supplementary Information (ESI) available: Experimental section of CQS synthesis and device fabrication; Simulated PbS CQDs solar cells' EQE curves; as-obtained PbS CQDs characterizations; the selection of electron blocking layer for 0.7 eV PbS CQDs solar cells; additional device performance tables. See DOI: 10.1039/x0xx00000x

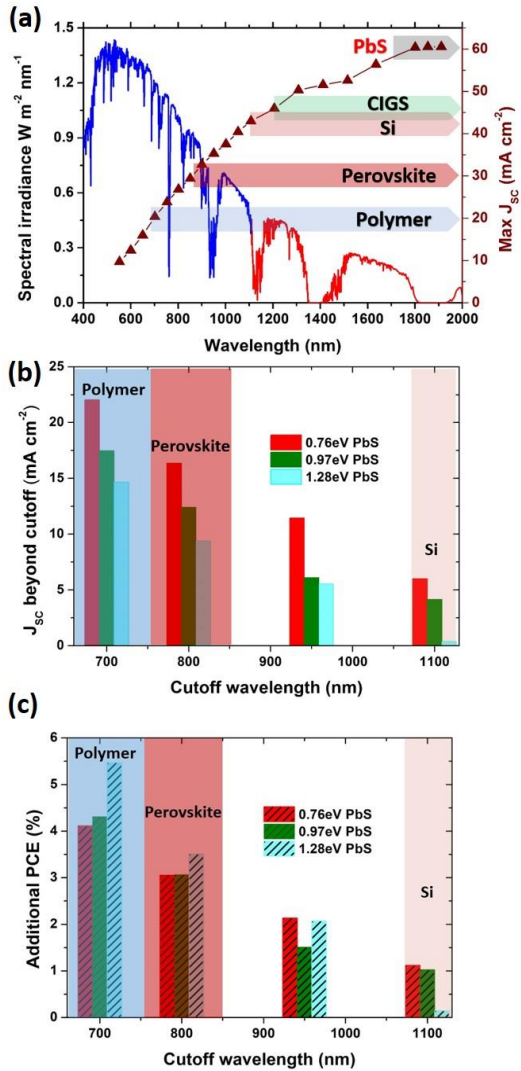

Figure 1 (a) Left axis: spectral irradiance for AM 1.5G spectrum, as well as the spectral non-response range for the common studied solar materials: polymer, perovskite, Si, CIGS, and PbS right axis: maximum short circuit current density (max $J_{SC}$) calculated from the AM1.5 G solar spectrum along with absorption edge (300 nm – absorption edge); (b) The $J_{SC}$ generated from the bottom cell in a mimic 4 terminal tandem cell using 0.76 eV, 0.97 eV and 1.28 eV PbS as a bottom subcell, respectively, and varied absorption edge (600-1100 nm) top cells. (c) The corresponding additional gain of power conversion efficiency from the PbS bottom cell (the FF of the PbS CQD cell was taken at 0.6 and the $V_{OC}$ value according to literature data for PbS CQD cells). (Note: cutoff wavelength or cutoff indicates the absorption edge of solar absorbers.)

CQD solar cells, respectively and a FF value of 60 % was assumed for all the devices.) An additional PCE of 5.5 % and 3.5 % could be achieved by simply placing 1.28 eV PbS CQDs solar cell at the back of a typical polymer and perovskite cell. The importance of SWIR PbS CQD cells is outlined when considering front cells of lower bandgap such as Silicon or CIGS based cells. In the case of Si, for example, an extra 1.1 % could be gained by using a 0.7 eV bandgap PbS CQD cell mounted at the back in a 4-terminal configuration.

Thus far, PbS CQD solar cells with an optimal bandgap for a single junction cell[23] of 1.3 eV have been extensively studied with efficiencies as high as 12 % been reported recently.[23-25] However, high efficiency low bandgap PbS(e) CQDs solar cells have been scarcely reported[26] with the best efficiency reported so far of 5.08 % by Wang et al using ZnO nanowires as the electron acceptor.[27] Here we report a planar low bandgap PbS CQD photovoltaic device with optimized electron acceptor and electron blocking layers that has reached a record high $J_{SC}$ of 37 mA/cm$^2$ under full solar spectrum and power conversion efficiency of 6.4 %.

## Results and discussion

The 0.7 eV PbS CQDs we used for the devices have been synthesized following a previously reported low temperature multi-injection method.[28, 29] This approach has led to PbS QDs with very narrow size dispersion as illustrated by the sharp absorption features (FWHM=90nm) in **Figure S2a** and a size dispersion of 8% (**Figure S2b, c**). The band diagram and device structure of the optimized devices are shown in Figure 2a. The device employs a chloride doped ZnO (Cl_ZnO) layer on top of which a 385 nm thick layer of PbS CQDs treated with a mixed ligand of $ZnI_2$/MPA [21, 22] and a thin layer of EDT treated PbS CQDs of higher bandgap. In **Figure 2b**, a cross-sectional FIB SEM image of a typical 0.7 eV PbS CQDs solar cell is presented, where a layer of Cl_ZnO atop ITO acted as the electron acceptor layer, PbS CQDs layer comprised of two parts, one was the $ZnI_2$/MPA treated 0.7 eV PbS CQDs layer as the active layer (385 nm), and the other is a thin EDT treated PbS CQD layer (around 50 nm) as the electron blocking layer, and the device was completed by a layer of Au electrode atop the electron blocking layer. The use of Cl_ZnO instead of neat ZnO has been instrumental in performance improvement in view of the more favorable band alignment compared to that of ZnO with the low bandgap PbS CQD absorber.[24, 30] **Table S1** compares the performance of devices using ZnO and Cl_ZnO as the electron acceptor. An additional important feature has been the selection of the EDT blocking layer.[31, 32] We have found that the use of a larger bandgap PbS CQD further improved the device performance by simultaneously increasing $J_{SC}$, $V_{OC}$ and FF (See Supporting Information **Figure S3** and **Table S2**), as a result of suppressed surface recombination when a larger PbS bandgap is used between the absorber and the metal back electrode.

In order to maximize the infrared solar harnessing capabilities of the device, an additional thickness optimization study of the active layer has been employed to tune the Fabry-Perot resonance to the solar spectrum band around 1400-1800 nm, achieving a nice solar spectrum matching profile (**Figure 2c**). This resulted in a dramatic EQE enhancement in this energy band reaching a value of 60 % at 1620 nm and led to a $J_{SC}$ of 37mA/cm$^2$. Device stability is also critical to photovoltaic applications. We studied the preliminary photostability of these low bandgap PbS QDs by exposing the non-encapsulated PbS CQD device in ambient conditions, continuously under AM1.5 solar illumination during the JV measurements. The evolution of the device performance as a function of time is shown in Figure S4a. Device performance progressively improves at the first hour due to the light soaking of the Cl_ZnO layer, and is stabilized for the next hour under continuous AM1.5 solar illumination. To further demonstrate the air stability of the devices, device performance was monitored during 90 days as a long term material stability test shown in Figure S4b. $V_{OC}$ remains constant for the whole test time, while $J_{SC}$ and FF slightly decrease with time in the first 17 days and finally stabilize, leading to slightly decreased PCE from 6.4% to 6.2%. Overall the solar cell performance retains 97% of its initial PCE after storage in ambient air without any device encapsulation for 90 days. The use of the back heterojunction between ZnI$_2$/MPA and EDT treated QDs may have led to this stability as the charge separation interface has been moved from a metal/CQD interface to a buried QD/QD interface.

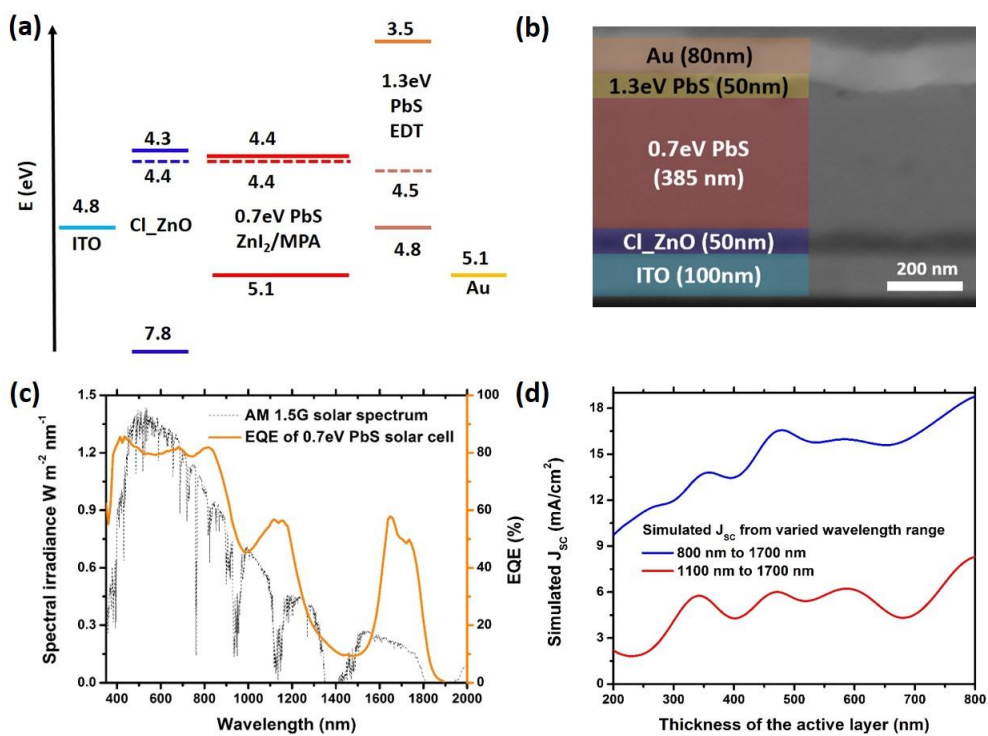

Figure 2(a) The energy level diagram of the PbS QDs solar cell referenced to the vacuum level, (detailed UPS data shown in Figure S3) (b) FIB cross section image of the typical device structure, (c) the AM1.5 spectrum plotted with the EQE of the SWIR PbS CQD solar cell; (d) Simulated $J_{SC}$ generated in the 0.7 eV PbS CQD solar cell as a function of the active layer thicknesses, when it forms a mimic 4 terminal tandem cells in combination with a subcell of perovskite (800 nm) or Si (1100 nm). (The blue curve corresponds to perovskite filter and the red one to Si filter).

To evaluate the potential of this optimized device as a subcell to harvest the NIR and SWIR parts of the solar spectrum in combination with other established PV technologies, we have measured the performance of the best solar matching cell by using a methyl ammonium lead iodide (MAPbI$_3$) perovskite filter and an intrinsic double polished 500 μm Si wafer between the device and the AM1.5 solar simulator. The device performance with and without optical filtering is summarized in **Table 1**. The infrared PV device can deliver an additional $J_{SC}$ of 5.5 mA/cm$^2$ when placed atop a polished 500 μm Si wafer and 13 mA/cm$^2$ when placed atop a 300 nm thick perovskite film, leading to an additional PCE of 0.71 % and 2 %, respectively. These experimentally achieved values of $J_{SC}$ are in very good agreement with the theoretically predicted values from optical simulations shown in **Figure 2d,** as a function of the PbS CQD active layer thickness.

Table 1. Device performance summary under AM1.5G solar simulator, perovskite filter and Si filter.

| Varied filters | $V_{OC}$ [V] | $J_{SC}$ [mA/cm$^2$] | FF [%] | PCE [%] |
|---|---|---|---|---|
| AM1.5 G (no filter) | 0.31 | 37.01 | 56 | 6.39 |
| Perovskite filter | 0.27 | 13.33 | 56 | 2.00 |
| Si filter | 0.23 | 5.50 | 56 | 0.71 |

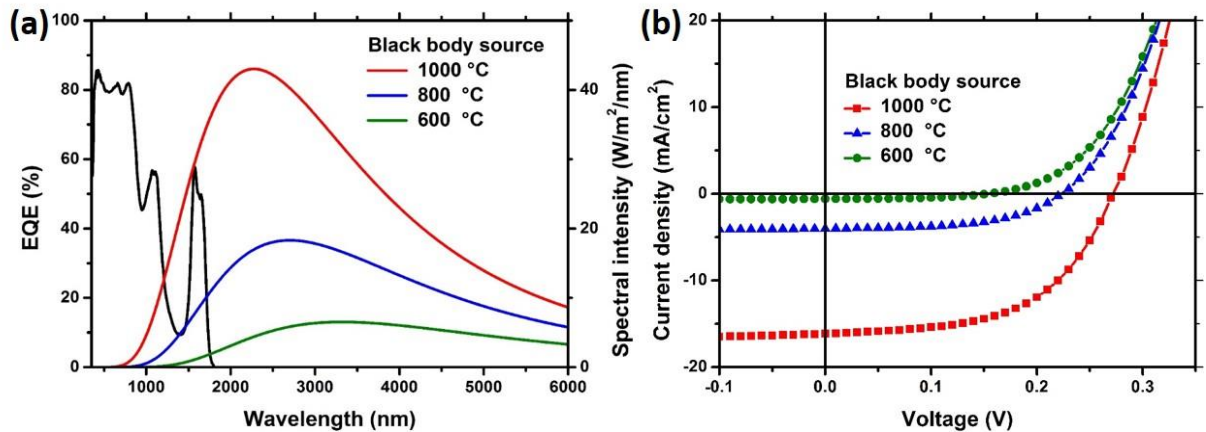

Figure 3 (a) Left axis: EQE curve of the best 0.7 eV PbS CQD solar cell; right axis: spectral radiation intensity for a black body source at 600, 800 and 1000 °C, calculated based on the Plank's equation and measured incident power density; (b) JV curves measured using the corresponding black body source as the illuminator.

**Table 2.** Thermo-photovoltaic performance under a black body source at varied temperature. The above band gap PCE were calculated based on the available incident power, indicated as above band gap power density (integrated from 100-1800 nm).

| Black body temperature (°C) | $V_{OC}$ [V] | $J_{SC}$ [mA/cm$^2$] | FF [%] | Above band gap power density [mW/cm$^2$] | Above band gap PCE [%] |
|---|---|---|---|---|---|
| 600 | 0.15 | 0.58 | 50 | 1.5 | 2.90 |
| 800 | 0.22 | 4.02 | 55 | 10.16 | 4.79 |
| 1000 | 0.27 | 16.13 | 55 | 40.15 | 5.97 |

In view of the device's spectral coverage down to 0.7 eV with EQE up to 60 % at these energies in **Figure 3a** left axis, we posited that a decent performance is within reach when the cell is used to harness infrared energy as a TPV device. We have tested our device under a black body source at 600, 800 and 1000 °C. The corresponding spectral radiation intensity in Figure 3a right axis is calculated by using the Plank's equation together with the measured value of the radiation power density received by the device. The total radiation power density impinging on the device was measured via a calibrated power meter PM200 (Thorlabs), and the above band gap power density at a certain temperature for the 0.7 eV bandgap PbS CQD device was integrated in the range of 100-1800 nm (**Table 2**). JV curves of the device under a black body source at various temperatures are shown in Figure 3b and their corresponding TPV performance is summarized in Table 2. The above band gap PCEs of 2.9 %, 4.8 % and 5.97 % were calculated based on the available incident above band gap photon energy (100-1800 nm) of 1.5, 10.16 and 40.15 mW/cm$^2$ respectively, corresponding to the black body source at 600, 800 and 1000 °C. The PCE of 4.8 % at 800 °C is notably higher than previously reported values of 2.7 % and constitutes a record performance for a solution processed TPV device.[33]

## Conclusions

In summary, we have demonstrated that a highly efficient solution processed low bandgap photovoltaic device based on 0.7 eV bandgap PbS CQDs with a $J_{SC}$ of 37 mA/cm$^2$ under full solar spectrum has been reached by optimizing the device structure, and 5.5 mA/cm$^2$ is achieved when the device placed at the back of a silicon wafer, resulting in PCE of 6.4 % and 0.7 % respectively. Moreover, this device has shown potential for thermal harvesting applications with an above bandgap PCE of ~6 % when operated as a thermo-photovoltaic cell under a 1000 °C blackbody radiator.

## Experimental

**Device characterization:** The current density- voltage (J-V) measurements were carried out using a Keithley 2400 source under AM1.5 illuminations (Oriel sol 3A, Newport Corporation). The accuracy of the measurements was determined as ±4 %. EQE spectra were recorded with a lock-in amplifier (Stanford Research System SR830) under chopped monochromatic light generated by white light source from a xenon lamp passing through a Newport Cornerstone 260 monochromator. The output power was calibrated with Newport 818-UV and Newport 838-IR photodetectors. All the devices were characterized in air under ambient conditions.

**UPS measurements:** UPS spectra of the PbS CQDs films were measured on ITO glass substrate. UPS measurements were performed with a SPECS PHOIBOS 150 hemispherical analyzer (SPECS GmbH, Berlin, Germany) in ultra-high vacuum conditions ($10^{-10}$ mbar). UPS measurements with monochromatic HeI UV source (21.2 eV).

**Optical simulations:** The optical simulation is based on the transfer matrix method. By keeping the thicknesses of other layers intact, except for the active layer thickness, $J_{SC}$ is generated with the variation of the active layer thickness.

**Black body source:** IR-508/301 Blackbody reference source from Infrared system development was used and the temperature was set to be 600, 800 and 1000 °C. The radiation power densities received by the device at each radiation temperature were measured to be 74.87, 166.54 and 326.46 mW/cm$^2$ via a Thorlabs PM200 power meter through an aperture with diameter of 2 mm at the position of the device. The detective range of the power meter is 0.2-10.6 µm.

**TPV performance measurements:** The J-V measurements were carried out using a Keithley 2400 source under the IR-508/301 Blackbody source at 600, 800 and 1000 °C. The corresponding above bandgap power densities of 1.5, 10.16 and 40.15 mW/cm$^2$ were obtained by integrating the blackbody spectral radiation density profile from 100 to 1800 nm taking into account of the total power above-mentioned.

## Conflicts of interest

There are no conflicts to declare.

## Acknowledgements

S.G., I.R. and S.P. contributed equally to this work. The authors acknowledge financial support from the European Research Council (ERC) under the European Union's Horizon 2020 research and innovation programme (grant agreement no. 725165), the Spanish Ministry of Economy and Competitiveness (MINECO), and the "Fondo Europeo de Desarrollo Regional" (FEDER) through grant TEC2017-88655-R. The authors also acknowledge financial support from Fundacio Privada Cellex, the program CERCA and from the Spanish Ministry of Economy and Competitiveness, through the "Severo Ochoa" Programme for Centres of Excellence in R&D (SEV-2015-0522). S.C. acknowledges support from a Marie Curie Standard European Fellowship ("NAROBAND", H2020-MSCA-IF-2016-750600). I.R. acknowledges support from the Ministerio de Economía, Industria y Competitividad of Spain via a Juan de la Cierva fellowship. We are also thankful to Dr. Q. Liu for providing us with a perovskite thin film layer.## References

1. C. H. Henry, *J Appl Phys*, 1980, **51**, 4494-4500.
2. A. D. Vos, *J Phys D: Appl Phys*, 1980, **13**, 839.
3. G. W. Charache, J. L. Egley, D. M. Depoy, L. R. Danielson, M. J. Freeman, R. J. Dziendziel, J. F. Moynihan, P. F. Baldasaro, B. C. Campbell, C. A. Wang, H. K. Choi, G. W. Turner, S. J. Wojtczuk, P. Colter, P. Sharps, M. Timmons, R. E. Fahey and K. Zhang, *J Electron Mater*, 1998, **27**, 1038-1042.
4. T. J. Coutts, *Sol Energ Mat and Sol C*, 2001, **66**, 443-452.
5. A. Datas, *Sol Energ Mat and Sol C*, 2015, **134**, 275-290.
6. A. Lenert, D. M. Bierman, Y. Nam, W. R. Chan, I. Celanović, M. Soljačić and E. N. Wang, *Nat Nanotech*, 2014, **9**, 126.
7. L. M. Fraas, J. E. Avery, V. S. Sundaram, V. T. Dinh, T. M. Davenport, J. W. Yerkes, J. M. Gee and K. A. Emery, 1990.
8. B.-C. Juang, R. B. Laghumavarapu, B. J. Foggo, P. J. Simmonds, A. Lin, B. Liang and D. L. Huffaker, *Appl Phys Lett*, 2015, **106**, 111101.
9. C. Ungaro, S. K. Gray and M. C. Gupta, *Opt Express*, 2015, **23**, A1149-A1156.
10. G. M. A., E. Keith, H. Yoshihiro, W. Wilhelm and D. E. D., *Prog Photovoltaics: Research and Applications*, 2012, **20**, 606-614.
11. J. Fernández, F. Dimroth, E. Oliva, M. Hermle and A. W. Bett, *AIP Conference Proceedings*, 2007, **890**, 190-197.
12. J. P. Mailoa, M. Lee, I. M. Peters, T. Buonassisi, A. Panchula and D. N. Weiss, *Energ Environ Sci*, 2016, **9**, 2644-2653.
13. T. Ming, J. Lian, W. Yuanyuan, D. Pan, W. Qingsong, L. Kuilong, Y. Ting, Y. Yao, L. Shulong and Y. Hui, *Appl Phys Express*, 2014, **7**, 096601.
14. M. W. Dashiell, J. F. Beausang, H. Ehsani, G. J. Nichols, D. M. Depoy, L. R. Danielson, P. Talamo, K. D. Rahner, E. J. Brown, S. R. Burger, P. M. Fourspring, W. F. Topper, P. F. Baldasaro, C. A. Wang, R. K. Huang, M. K. Connors, G. W. Turner, Z. A. Shellenbarger, G. Taylor, J. Li, R. Martinelli, D. Donetski, S. Anikeev, G. L. Belenky and S. Luryi, *IEEE Transactions on Electron Devices*, 2006, **53**, 2879-2891.
15. J. Tang and E. H. Sargent, *Adv Mater*, 2011, **23**, 12-29.
16. E. H. Sargent, *Nat Photon*, 2009, **3**, 325.
17. A. G. Pattantyus-Abraham, I. J. Kramer, A. R. Barkhouse, X. Wang, G. Konstantatos, R. Debnath, L. Levina, I. Raabe, M. K. Nazeeruddin, M. Gratzel and E. H. Sargent, *ACS Nano*, 2010, **4**, 3374-3380.
18. G. F. Burkhard, E. T. Hoke and M. D. McGehee, *Adv Mater*, 2010, **22**, 3293-3297.
19. E. Centurioni, *Appl. Opt.*, 2005, **44**, 7532-7539.
20. L. A. A. Pettersson, L. S. Roman and O. Inganäs, *J. Appl. Phys.*, 1999, **86**, 487-496.
21. S. Pradhan, A. Stavrinadis, S. Gupta, Y. Bi, F. Di Stasio and G. Konstantatos, *Small*, 2017, **13**.
22. B. Yu, P. Santanu, G. Shuchi, A. M. Zafer, S. Alexandros and K. Gerasimos, *Adv Mater*, 2018, **30**, 1704928.
23. M. Liu, O. Voznyy, R. Sabatini, F. P. Garcia de Arquer, R. Munir, A. H. Balawi, X. Lan, F. Fan, G. Walters, A. R. Kirmani, S. Hoogland, F. Laquai, A. Amassian and E. H. Sargent, *Nat Mater*, 2017, **16**, 258-263.
24. J. Choi, Y. Kim, J. W. Jo, J. Kim, B. Sun, G. Walters, F. P. Garcia de Arquer, R. Quintero-Bermudez, Y. Li, C. S. Tan, L. N. Quan, A. P. T. Kam, S. Hoogland, Z. Lu, O. Voznyy and E. H. Sargent, *Adv Mater*, 2017, **29**, 1702350.
25. J. Xu, O. Voznyy, M. Liu, A. R. Kirmani, G. Walters, R. Munir, M. Abdelsamie, A. H. Proppe, A. Sarkar, F. P. García de Arquer, M. Wei, B. Sun, M. Liu, O. Ouellette, R. Quintero-Bermudez, J. Li, J. Fan, L. Quan, P. Todorovic, H. Tan, S. Hoogland, S. O. Kelley, M. Stefik, A. Amassian and E. H. Sargent, *Nat Nanotech*, 2018, **13**, 456-462.
26. O. E. Semonin, J. M. Luther, S. Choi, H.-Y. Chen, J. Gao, A. J. Nozik and M. C. Beard, *Science*, 2011, **334**, 1530-1533.
27. H. Wang, T. Kubo, J. Nakazaki and H. Segawa, *ACS Energy Lett*, 2017, **2**, 2110-2117.
28. S. Goossens, G. Navickaite, C. Monasterio, S. Gupta, J. J. Piqueras, R. Pérez, G. Burwell, I. Nikitskiy, T. Lasanta, T. Galán, E. Puma, A. Centeno, A. Pesquera, A. Zurutuza, G. Konstantatos and F. Koppens, *Nat Photon*, 2017, **11**, 366.
29. J. W. Lee, D. Y. Kim, S. Baek, H. Yu and F. So, *Small*, 2016, **12**, 1328-1333.
30. J. Choi, J. W. Jo, F. P. G. Arquer, Y.-B. Zhao, B. Sun, J. Kim, M.-J. Choi, S.-W. Baek, A. H. Proppe, A. Seifitokaldani, D.-H.

Nam, P. Li, O. Ouellette, Y. Kim, O. Voznyy, S. Hoogland, S. O. Kelley, Z.-H. Lu and E. H. Sargent, *Advanced Materials*, 2018, **30**, 1801720.
31. C.-H. M. Chuang, P. R. Brown, V. Bulović and M. G. Bawendi, *Nat Mater*, 2014, **13**, 796-801.
32. N. Zhang, D. C. J. Neo, Y. Tazawa, X. Li, H. E. Assender, R. G. Compton and A. A. R. Watt, *ACS Appl Mater Interfaces*, 2016, **8**, 21417-21422.
33. A. Kiani, H. Fayaz Movahed, S. Hoogland, O. Voznyy, R. Wolowiec, L. Levina, F. P. Garcia de Arquer, P. Pietsch, X. Wang, P. Maraghechi and E. H. Sargent, *ACS Energy Lett*, 2016, **1**, 740-746.